\documentclass[iop,apj]{emulateapj}
\usepackage{amsmath,amssymb}
\usepackage{graphicx}
\usepackage{CJK}
\usepackage{natbib}
\usepackage{color}
\usepackage[
    breaklinks,
    colorlinks=true,
    urlcolor=blue,
    linkcolor=red,
    citecolor=blue
    ]{hyperref}
\graphicspath{{./}{figs/}}
\def\GALEX{{\it GALEX}}
\newcommand{\angstrom}{{\rm \mathring A}}
\usepackage[english]{babel}
\usepackage{blindtext}

\newcommand{\lya}{\hbox{Ly$\alpha$}}
\newcommand{\ciiii}{\hbox{C\,{\sc iv}}}
\newcommand{\ciii}{\hbox{C\,{\sc iii]}}}
\newcommand{\mgii}{\hbox{Mg\,{\sc ii}}}

\begin{document}
\begin{CJK*}{UTF8}{gbsn}
\title{The Timescale-Dependent Color Variability of Quasars Viewed with \GALEX}
\author{
    Fei-Fan Zhu (朱飞凡),
    Jun-Xian Wang (王俊贤),
    Zhen-Yi Cai (蔡振翼),
    Yu-Han Sun (孙玉涵)}
\affil{CAS Key Laboratory for Researches in Galaxies and Cosmology,
University of Science and Technology of China, Chinese Academy of Sciences, 
Hefei, Anhui 230026, China}
\email{zff1991@mail.ustc.edu.cn, jxw@ustc.edu.cn}
\begin{abstract}
In recent work done by \citeauthor{2014ApJ...792...54S}, the color variation of quasars, namely the bluer-when-brighter trend, was found to be timescale-dependent using SDSS $g/r$ band light curves in the Stripe 82. 
Such timescale dependence, i.e., bluer variation at shorter timescales, supports the thermal fluctuation origin of the UV/optical variation in quasars, and can be well modeled with the inhomogeneous accretion disk model. In this paper, we extend the study to much shorter wavelengths in the rest frame (down to extreme UV), using {\it GALaxy Evolution eXplorer} (\GALEX) photometric data of quasars collected in two ultraviolet bands (near-UV and far-UV). 
We develop Monte-Carlo simulations to correct possible biases due to the considerably larger photometric uncertainties in \GALEX{} light curves (particularly in far-UV, comparing with SDSS $g/r$ bands), which otherwise could produce artificial results. 
We securely confirm the previously discovered timescale dependence of the color variability  with independent datasets and at shorter wavelengths.
We further find the slope of the correlation between the amplitude of color variation and timescale however appears even steeper than that predicted by the inhomogeneous disk model, which assumes that disk fluctuations follow damped random walk process. 
In line with the much flatter structure function observed in far-UV comparing with that at longer wavelengths, this implies deviation from DRW process in the inner disk where rest frame extreme UV radiation is produced. 
\end{abstract}

\section{Introduction}\label{sect:intro}
As a defining feature of quasars and active galactic nuclei (AGNs), variability starts to gain more attention because it holds otherwise inaccessible information of them. 
The energy source of these shinning sources is widely accepted to be dominated by the thermal radiation from the accretion disk \citep{1973A&A....24..337S}. 
As suggested by the reverberation mapping projects \citep{2004ApJ...613..682P}, variability should be traced back to the inner parts of AGNs, including the accretion disk, which contributes optical and UV photons, and the presumed corona, which dominates over the X-ray band. 

The corona is generally assumed to work as a light bulb above the disk and modulates radiation from the disk, encoding information about sizes and distances in the form of time lags between light curves in different photometric bands. 
This is the famous X-ray reprocessing model \citep{1991ApJ...371..541K}, and has been tested in great details with nearby Seyferts \citep{2005ApJ...622..129S,2014MNRAS.444.1469M,2015ApJ...806..129E,2016ApJ...821...56F,2016MNRAS.456.4040T}. 
The correlation analysis of inter-bands (X-ray/UV/optical) light curves typically results in lags less than a few days (for a short review, see \citealt{2012MNRAS.423..451L}), likely corresponding to light travel time. 
However, it should be kept in mind that X-ray only contributes to a small fraction of AGNs' total bolometric luminosity, especially for brighter ones \citep{2005AJ....130..387S,2010A&A...512A..34L,2010ApJS..187...64G}, thus could be insufficient to produce the observed UV/optical variation \citep{2008RMxAC..32....1G}.

Different mechanisms are involved to explain the observed UV/optical variation in AGNs. 
These include changes in global accretion rates \citep{2006ApJ...642...87P,2008MNRAS.387L..41L,2011ApJ...731...50S,2012ApJ...758..104Z,2013A&A...554A..51G}, and the instability of the accretion disk \citep{1998ApJ...504..671K,2003MNRAS.342.1222C,2013A&A...560A.104M} with large temperature fluctuations \citep{2011ApJ...727L..24D,2012ApJ...744..147S,2014ApJ...783..105R,2014ApJ...792...54S}. 

Recent progresses on the UV/optical variability of AGNs show that it can be modeled by damped random walk (DRW) process \citep{2009ApJ...698..895K,2010ApJ...708..927K,2010ApJ...721.1014M,2013ApJ...765..106Z} or even more complicated ARMA (autoregressive moving average) model \citep{2014ApJ...788...33K}. Furthermore, it is revealed that the variability behavior is wavelength dependent in the sense that the variation in bluer bands is stronger than that in redder ones. As a result, AGNs appear bluer when they get brighter. Such bluer-when-brighter (BWB) trend has been confirmed for both nearby AGNs \citep{2010ApJ...711..461S} and quasars \citep{1985ApJ...296..423C,1990ApJ...354..446W,1991ApJ...366...64C,1999MNRAS.306..637G,2000ApJ...540..652W,2001ApJ...551..103T,2002ApJ...564..624T,2004ApJ...601..692V,2005ApJ...633..638W,2011A&A...525A..37M,2011A&A...527A..15W,2011ApJ...731...50S,2012ApJ...744..147S,2012ApJ...758..104Z,2012ApJ...759...88B,2014ApJ...783..105R,2014ApJ...792...54S,2016ApJ...822...26G}. 

The two aforementioned disk-related mechanisms for the origin of UV/optical variability can both explain such BWB trend. A third explanation involves the contamination from the host galaxy or other more stable components \citep{2003MNRAS.344..492H,2010ApJ...711..461S}. 
Using SDSS photometric monitoring of 9258 spectroscopic confirmed quasars in Stripe 82, \citet{2014ApJ...792...54S} discovered that the color variability (the BWB trend) is more prominent on shorter timescales than on longer ones, which was coined as timescale-dependent color variability. 
Given the fact that neither the changing accretion rate nor contamination from host galaxy can produce timescale-dependent behavior, an inhomogeneous disk with temperature fluctuations should step in. 

Based on the original model proposed by \citet{2011ApJ...727L..24D}, \citet[][hereafter Cai16]{2016ApJ...826....7C} developed a revised inhomogeneous accretion disk model, and found that such model can well explain the observed  timescale-dependent color variability in \citet{2014ApJ...792...54S}, primarily focusing on the slope of the relation between the amplitude of color variation and the timescale.
The underlying physics is the inner and hotter zones of the accretion disk fluctuate at shorter timescales, and thus produce faster and bluer variations.
Studying the variation of quasars at different timescales therefore provides an approach to probe the accretion disk in a spatially resolved manner.

The UV light curves of quasars recorded by \GALEX{} enable us to extend the study of \citet{2014ApJ...792...54S}  to shorter wavelengths, and probe the fluctuation at the inner most accretion disk. 
Using quasar light curves from \GALEX{} GR5, \citet{2011A&A...527A..15W} presented the ensemble near-UV (NUV) and far-UV (FUV) structure functions of quasars in the observed frame. 
They demonstrated that variation in FUV is stronger than that in NUV and they both triumph over the amplitudes of optical variability, also supporting the BWB diagram.  
Is such BWB trend revealed by \GALEX{} similarly timescale-dependent? 
To address this question, the contents of this work are orchestrated as follows: Section~\ref{sect:data} describes the data collected from \GALEX{} archive and Section~\ref{sect:method} makes use of the method introduced by \citet{2014ApJ...792...54S} to check the timescale dependence of color variability. 
We also present Monte-Carlo simulations to correct possible bias due to the large photometric uncertainties in the light curves. 
In Section~\ref{sect:diss}, we give the timescale-dependent color variation in different redshift bins and comparisons with the inhomogeneous model developed by Cai16.
Conclusions are  listed in Section~\ref{sect:conc}.
 
\section{Data description}\label{sect:data}

The {\it GALaxy Evolution eXplorer} (\GALEX{}) is a spaceborne telescope working at NUV and FUV bands, offering both imaging (for almost the whole sky) and spectroscopic observations. 
In this paper we only make use of the photometric data in NUV ($1770 \sim 2830 \angstrom$, centered at 2316 $\angstrom$; spatial resolution of 4.3 arcsec) and FUV ($1350 \sim 1785 \angstrom$, centered at 1539 $\angstrom$; spatial resolution of 5.2 arcsec).

\begin{figure}[!t]
\centering
\includegraphics[width=0.45\textwidth]{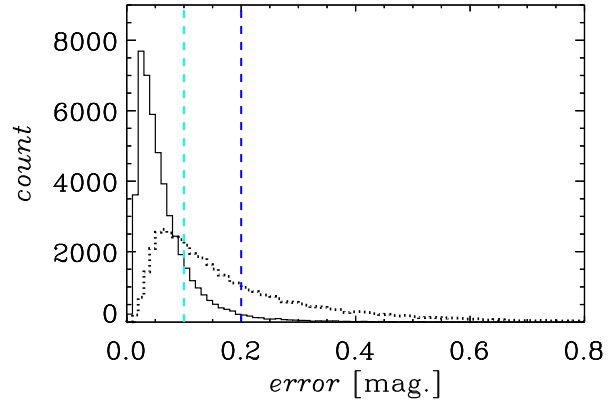}
\caption{The photometric error distribution for both NUV (the solid black histogram) and FUV (the dotted black one) bands of the matched SDSS quasars. Two vertical dashed lines, locating at 0.1 (NUV) and 0.2 (FUV) magnitude respectively, indicate the photometric uncertainty cuts we adopted, i.e., only photometric measurements with smaller errors are included.}
\label{fig:errorhistogram}
\end{figure}

After surveying the sky for almost a decade, \GALEX{} has accumulated more than 200 million photometric measurements \citep{2014Ap&SS.354..103B}. 
We cross-match \GALEX{} data release 6/7 (GR6/7) with the SDSS DR7 spectroscopically confirmed quasar catalog \citep{2010AJ....139.2360S}, whose spectroscopic properties have been measured by \citet{2011ApJS..194...45S}, resulting in a preliminary sample of 83228 quasars, observed in at least one \GALEX{} band.
A match is deemed positive if the matching radius is no more than 5 arcsec, which is similar to the spatial resolution of \GALEX{} observations. 
To compile a list of light curves suitable for variability study, we further reject observation epochs with \GALEX{} exposure times less than 200 seconds or those when targets fall on the edge of the detector (more than $0.55^{\circ}$ off the center, as suggested by \citet{2008AJ....136..259W}), since under both circumstances the photometry results are generally unreliable. 
Observations made in survey types other than all/medium/deep sky imaging survey (AIS, MIS, DIS) are excluded, since such surveys need special pipelines to account for contamination from extended galaxies or crowded neighbors to acquire reliable photometries. 
It is also useful to set constraints on photometric errors to avoid outliers. 
To be specific, we adopt a photometric uncertainty cut of 0.1 magnitude for NUV and 0.2 for FUV (see Fig.~\ref{fig:errorhistogram}). 
In the end, we require simultaneous photometric measurements in both bands for at least two epochs so as to calculate color variability. 
After all these steps, a catalog of 5282 quasars are chosen and used for the following analyses. 
The average number of observation epochs for the final sample is about 5.

\section{Methods and Result}\label{sect:method}

For the light curve of each quasar we obtain, any two photometry points of it ($m_{i}^{\rm NUV},m_{i}^{\rm FUV},t_i$ and $m_{j}^{\rm NUV},m_{j}^{\rm FUV},t_j$) form a data pair with its timescale defined as $\tau = \mid t_i - t_j \mid $. 
In the bottom panel of Fig.~\ref{fig:figsec2} we plot the number of such pairs as a function of the observed timescale.
Note that we do not use the rest frame timescale here, since for quasars over a large range of redshifts but monitored at the same observed timescale, the ensemble structure function would suffer from a systemic bias that at longest rest frame timescale, the data are dominated by lowest-z sources, and at shortest rest frame times scale by highest-z sources. The situation would be more complicated considering the significant gaps in the timescale of sampling (see the lowest panel of Fig. 2). Analyzing in the observed frame is however not affected by such bias. 

\subsection{Ensemble structure function}\label{subsect:SF}

\begin{figure}[!t]
\centering
\includegraphics[width=0.45\textwidth]{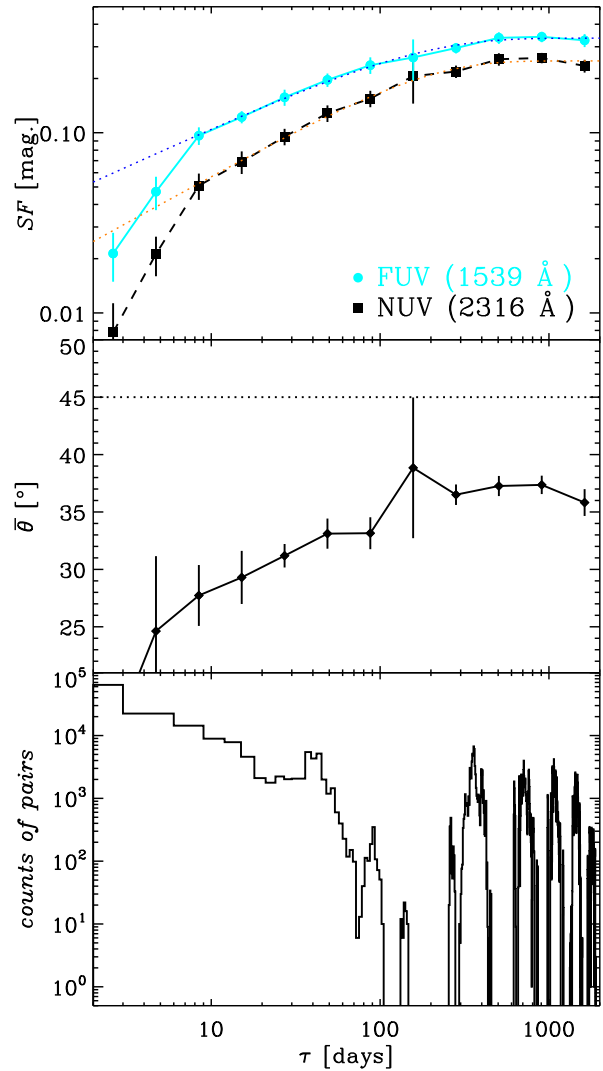}
\caption{Upper panel: ensemble structure functions calculated for the UV selected quasar sample in the observed frame (red squares for NUV and blue filled circles for FUV). Middle panel:  $\theta_{\textnormal{SF-ratio}}$'s timescale-dependent trend. The $45^\circ$ dashed horizontal line corresponds to no  color variation.
 Bottom panel: the histogram of data pairs as a function of time differences in observed frame. The abnormally large error bars at timescale of about 150 days in the two upper panels are due to the small number of data pairs. All error bars in the plot represent the 1 $\sigma$ bootstrapped uncertainty and the same goes for the rest figures.}
\label{fig:figsec2}
\end{figure}

We calculate the structure function following \citet{1996ApJ...463..466D}. 
Taking all the data pairs contributing to the variability at a certain timescale $\tau$, the ensemble structure function of a quasar sample is defined as
\begin{equation}
SF(\tau)=\sqrt{\dfrac{\pi}{2} \langle \mid m_i - m_j \mid \rangle ^{2} - 
\langle \sigma_{i} ^{2} + \sigma_{j} ^{2} \rangle} ,
\end{equation}
where $\sigma_i$ and $\sigma_j$ are photometric errors corresponding to magnitudes $m_i$ and $m_j$. 
As stated by \citet{2004ApJ...601..692V}, such form is more robust against the presence of outliers in the data than averaging the square of the magnitude differences. 
This equation conforms to the requirement of structure functions brought up in \citet{2016ApJ...826..118K} and can accurately subtract the noise term.
To estimate the errors of ensemble structure function, we have bootstrapped the quasar sample for 1000 times to recalculate the ensemble structure functions and take their standard deviations as the uncertainties of the structure functions. 
The abnormally larger error bar at timescale bin of $\simeq$ 150 days is due to few data pairs (as shown in the bottom panel of Fig.~\ref{fig:figsec2}).
Aso not that due to the contamination of delayed varying emission lines, the calculated structure functions could be biased. We will dig into their effect on analysis of variability in Section~\ref{sect:redshift}.

The ensemble structure functions for NUV and FUV bands are shown in the top frame of Fig.~\ref{fig:figsec2}, broadly consistent with those of \citet{2011A&A...527A..15W}, except that the structure functions presented in \citet{2011A&A...527A..15W} were calculated for a quasar sample selected to have significant variation. 
Thus their variation amplitudes are slightly larger than ours.
Nevertheless, the structure functions of both work show variability amplitude increases as the timescale prolongs, with significantly stronger variation in FUV than in NUV.  
At shorter timescales (less than 10 days), quasars typically vary less than 0.1 magnitude, while for much longer timescales, variability amplitudes can be as high as 0.3 magnitude for FUV. 
Both structure functions tend to flatten at timescales longer than about 300 days. 

\citet{2016ApJ...826..118K} inspired us to fit the two structure functions using his Equation~19 introduced in the paper, to investigate possible deviation from DRW model. 
We slightly modify the equation by neglecting the noise term (as it has been subtracted off in our structure functions) and result in a three-parameter structure function model. The three parameters include the power index $\beta$ ($\beta=1$ for DRW process), de-correlation timescale $\tau_{c}$ and variance at long timescale $SF_{\infty}$ :
\begin{equation}
SF_{}(\tau) = SF_{\infty} \sqrt{1 - \rm{exp}(-\dfrac{\tau}{\tau_c})^{\beta}}.
\end{equation}
We fit our results omitting data points less than 6 days, where the structure functions show rapid drops toward shorter timescales, and would correspond to a much steeper PSD, significantly deviating from the DRW model. 
We note that steeper PSDs on shorter timescales (hours to months) has also been revealed in \citet{2011ApJ...743L..12M} using \emph{Kepler} light curves of AGNs. 
Out titting shows the $SF_{\infty}$ is $0.25\pm0.01$ mag for NUV and $0.33\pm0.02$ mag for FUV. 
And the deconvolution timescale $\tau_{c}$ is $167\pm46$ days for NUV and $142\pm46$ days for FUV, shorter than the timescale measured for optical bands \citep[e.g., $354\pm168$ days for $r$ band in ][]{2016ApJ...826..118K}. 
The power index for NUV is consistent with DRW model's prediction, measured to be $1.04\pm0.13$, while for FUV, the index is $0.84\pm0.11$. 
Slopes of the structure functions on timescales shorter than $\tau_{c}$ are half of $\beta$, namely, $0.52\pm0.06$ for NUV and $0.42\pm0.05$ for FUV. 
Be aware that these parameters were obtained in the observed frame rather than the rest frame, and special care need to be taken when comparing them with those of other works. 
Implications of our values will be discussed in Section~\ref{sect:comp}. 

Note that the stronger variation in FUV than in NUV seen in the structure functions indeed demonstrates a BWB trend.
To inspect the difference between FUV and NUV at various timescales, we introduce $\theta_{\textnormal{SF-ratio}}$ as
\begin{equation}
\theta_{\textnormal{SF-ratio}}(\tau) = \arctan \dfrac{SF_{\rm NUV}(\tau)}{SF_{\rm FUV}(\tau)}.
\end{equation}
This value,  similar to the $\theta$ defined by \cite{2014ApJ...792...54S}, can also be used to quantify the amplitude of the BWB trend. If $\theta_{\textnormal{SF-ratio}}$ equals $45^{\circ}$, the BWB trend vanishes, and the more it deviates from $45^{\circ}$, the more intense is the color variability (i.e., bluer when brighter if $<$ $45^{\circ}$, and redder when brighter if $>$ $45^{\circ}$). The middle panel of Fig.~\ref{fig:figsec2}
plots the dependence of $\theta_{\textnormal{SF-ratio}}$ on timescale, indicating stronger BWB trend at shorter timescales, similar to that reported by \cite{2014ApJ...792...54S}.
Logically, such approach does not require simultaneous light curves in two bands as we do.
We note that although the ratio of the structure functions between two bands also reflect the timescale dependence of the color variation in quasars, during this process any information of simultaneous observations in two bands were lost. 
Below we will derive $\theta$ following the direct approach raised by \cite{2014ApJ...792...54S} and compare with $\theta_{\textnormal{SF-ratio}}$.

\subsection{Timescale-dependent color variability of quasars}\label{subsect:Color}
The procedures to calculate $\theta$ and the interpretation on it have been detailed in \citet{2014ApJ...792...54S}. Here we will briefly introduce the relevant equations used. 
Individual $\theta$ for all observational data pairs is defined as
\begin{equation}
    \theta(\tau)=\arctan \left(\frac{m^{\rm NUV}(t+\tau)-m^{\rm NUV}(t)}{m^{\rm FUV}(t+\tau)-m^{\rm FUV}(t)}\right).
\end{equation}
Some of them need to be transformed so as to fall into the range of [$-45^{\circ}$, $135^{\circ}$]. 
In this way they can be statistically averaged over a certain timescale bin to indicate color variability:
\begin{equation}
\bar\theta(\tau)=\dfrac{\sum\limits_i^N \theta_i(\tau)}{N}.
\end{equation}
Here $N$ represents the number of data pairs for the timescale bin.
The derived $\theta(\tau)$ is plotted in the upper panel of Fig.~\ref{fig:figsec3}. 
In \citet{2014ApJ...792...54S}, only data pairs with variation $>$ 3$\sigma$ between two epochs were adopted to avoid possible bias induced by the photometric uncertainties. 
In this work, due to the larger photometric errors of $GALEX$ light curves comparing with SDSS (see Fig.~\ref{fig:errorhistogram}), we only exclude data pairs with variation $<$ 1$\sigma$ in order to keep sufficient number of pairs, and perform Monte-Carlo simulations to correct the possible bias owing to the large photometric errors.

We note that $\theta$ can be easily transformed into the ratio  of magnitude variation in two bands ($\Delta$m$_{FUV}$/$\Delta$m$_{NUV}$ in this work), and such quantity measures the BWB trend equally \citep[e.g.][]{2012ApJ...744..147S}. However, as we shown in \citet{2014ApJ...792...54S},  $\theta$ behaves better as it spans a limited range and can be easily averaged, while $\Delta$m$_{FUV}$/$\Delta$m$_{NUV}$ can reach infinity due to photometric noise. Nevertheless, we plot as well the $\Delta$m$_{FUV}$/$\Delta$m$_{NUV}$ derived from the averaged $\theta$ in the corresponding figures. 

\subsection{Simulations to retrieve the intrinsic color variation}

The bias to the measurement of $\theta$ due to photometric uncertainties is nontrivial, especially when the photometric errors are comparable to or even larger than the amplitude of intrinsic variation. 
In case of no intrinsic variation, $\theta$ would be determined by the ratio of the independent photometric errors in two bands. 
As pointed out in \citet{2014ApJ...792...54S}, if the photometric uncertainties in the bluer band are larger than those in the redder band, the photometric errors alone can produce artificial $\theta$ smaller than $45^{\circ}$ even there is no intrinsic BWB trend. 
The effect of photometric errors is timescale-dependent, i.e., weaker at longer timescales, since the intrinsic variations are much stronger at longer timescales. 
While such effect could be negligible for \citet{2014ApJ...792...54S}, as SDSS $g$ and $r$ bands have rather small and comparable photometric errors, and data pairs with flux differences not dominated by intrinsic variations ($<$ 3$\sigma$) are excluded, it is particularly important for this study as the photometric errors in $GALEX$ bands are considerably larger than SDSS $g$ and $r$ bands, and more importantly larger in FUV than in NUV (see Fig.~\ref{fig:errorhistogram} and Fig.~2 in \citet{2014ApJ...792...54S}).

To rectify such bias effect, we perform Monte-Carlo simulations to recover the intrinsic value of $\theta$. 
We start from the observed FUV structure function to simulate FUV variations at various timescales. 
The FUV magnitude differences follow a normal distribution with zero mean and variance equaling to the observed FUV structure function value at given timescale. 
Once assuming the intrinsic color variability quantified by a constant input $\theta_{\rm int}$,  the NUV variations are given by 
\begin{equation}
\Delta{m}_{\rm NUV}(\tau) = \Delta{m}_{\rm FUV}(\tau)\tan{\theta_{\rm int}}.
\end{equation}
This method rests upon the assumption that variations in two bands occur in phase, which is supported by the short time delays (hours or even shorter for FUV and NUV) between bands for nearby Seyfert galaxies, as mentioned in the introduction.

We further add randomized Gaussian errors to $\Delta{m}_{\rm NUV}$ and $\Delta{m}_{\rm FUV}$, respectively, with the variance of the errors randomly drawn from the real observed photometric uncertainties within the corresponding timescale bins. 
Note that a factor of $\sqrt{2}$ needs to be considered to account for error propagation of magnitude differences. 

\begin{figure}[!t]
\centering
\includegraphics[width=0.45\textwidth]{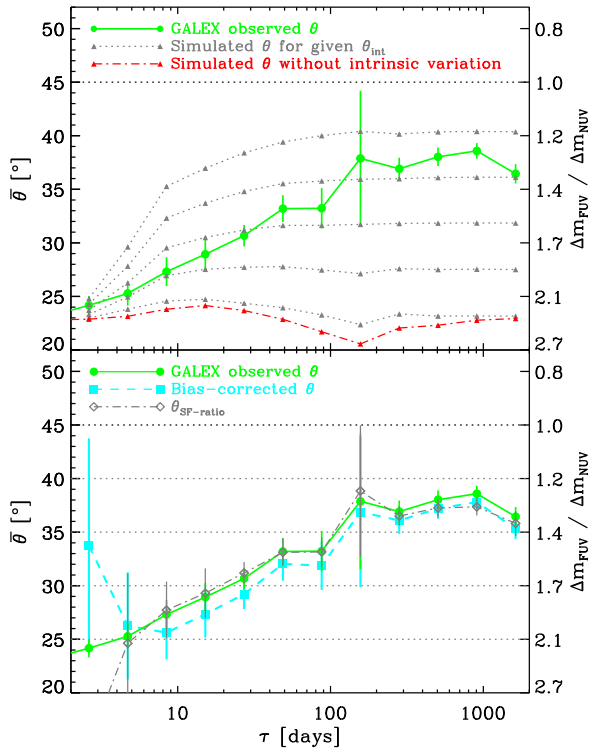}
\caption{This figure shows how the correction is done to the \GALEX{} observed $\theta$. 
The green solid line with filled circles in both panels indicates the $\theta$ directly calculated from \GALEX{} observed data. 
The gray dotted lines with filled triangles are the simulated $\theta$s correspond to intrinsic $\theta$s fixed at 40, 35, 30, 25, 20 from up to down, respectively, while the red dash-dot one represents the simulation without intrinsic variation at all. In the lower panel, the blue dashed line with filled squares is the bias-corrected $\theta$ and the thinner grey dash-dot line is $\theta_{\textnormal{SF-ratio}}$ introduced in Section~\ref{subsect:SF}.
$\theta$ can be effectively transformed into $\Delta$m$_{FUV}$/$\Delta$m$_{NUV}$, which are also labeled.
}
\label{fig:figsec3}
\end{figure}

Based on the simulated magnitude differences, we plot the output $\theta$ versus timescale for different input values in the upper panel of Fig.~\ref{fig:figsec3}.
We see that for constant input $\theta$s ($\geqslant 20^{\circ}$), the simulated output $\theta$s clearly show artificial timescale dependence, particularly at very short timescales. 
This is the direct reflection of the bias effect caused by large photometric errors.
We also notice that the simulated $\theta$ can be larger than the input intrinsic one, contrary to previous inference. 
This is the bias effect introduced by the averaging range we adopted([$-45^{\circ}$, $135^{\circ}$]), which would drag the mean $\theta$ values towards $45^{\circ}$.

The directly calculated $\theta-\tau$ relation is over-plotted as well, however showing steeper slope comparing with the simulated relations. 
This indicates that the artificial effect of photometric uncertainties alone can not explain the observed $\theta-\tau$ relation. 
Using the simulated relations between input $\theta_{\rm int}$ and output $\theta$ 
demonstrated as dotted lines in the top panel of Fig.~\ref{fig:figsec3}, the bias-corrected $\theta$ corresponding to the \GALEX{} observed $\theta$ can be easily retrieved.

In the lower panel of Fig.~\ref{fig:figsec3}, the bias-corrected $\theta-\tau$ is over-plotted as blue squares connected by a blue dashed line.
We see no strong difference between the bias-corrected $\theta-\tau$ and the observed one, except for that the former is slightly smaller and has large uncertainties especially at very short timescales.
This is because such bias is only dominant at very short timescales where intrinsic variation is too weak, and at short timescales the intrinsic color variation can not be well constrained due to the large noises. 
We emphasize that the small difference between bias-corrected $\theta-\tau$ and the directly observed one does not mean no correction is needed for future studies. 
The correction depends on the intrinsic $\theta-\tau$ relation, the amplitudes of the intrinsic variation, and the level of photometric uncertainties. 
The $\theta_{\textnormal{SF-ratio}}-\tau$ relation derived from structure functions appears to be consistent with both the bias-corrected and direct observed ones, confirming that we can use the ratio of structure functions to probe the timescale dependence of color variations.

\section{Discussion}\label{sect:diss}
\subsection{Redshift dependence}\label{sect:redshift}
\begin{figure}[!t]
\centering
\includegraphics[width=0.45\textwidth]{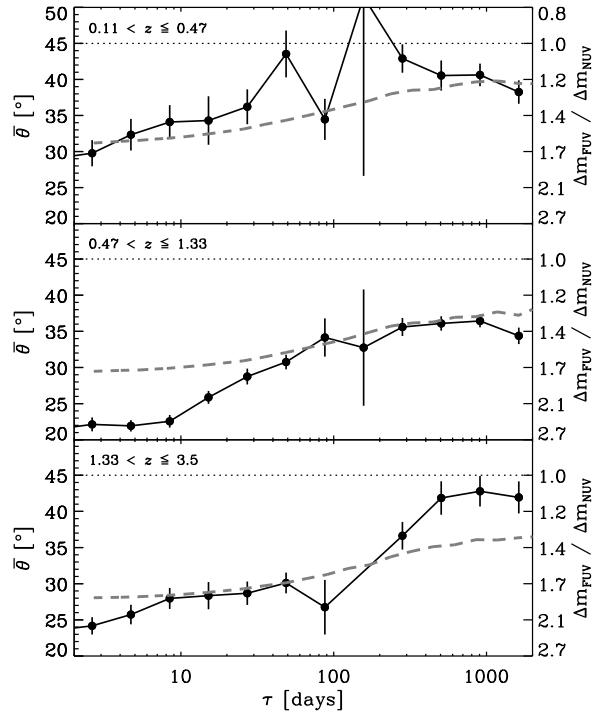}
\caption{
The color variations in observed frame for the three redshift bins (solid lines) are compared with the predictions of the revised inhomogeneous disk model given by Cai16 (grey dashed lines). 
In the lowest panel, the data point at $\sim$ 150 days was dropped as there are  too few data pairs in this bin.
Both $\theta$ and the corresponding $\Delta$m$_{FUV}$/$\Delta$m$_{NUV}$ are labeled.
}
\label{fig:figsec42}
\end{figure}

As having been discussed by \cite{2011A&A...527A..15W}, \cite{2014ApJ...792...54S} and \cite{2014ApJ...783...46K}, the presence of emission lines would differently affect the photometries of the broad bands considering the various redshifts of quasars, and further interfere the calculated properties of color variability. 
In the UV bandpass of quasar spectrum, the \lya{} emission line around 1216 $\angstrom$ is the most prominent one, in comparison with other possible contamination lines such as \ciiii, \ciii{} and \mgii. 
We follow the tackling method used in \cite{2011A&A...527A..15W} and select three redshift bins based on into which bandpass \lya{} would be redshifted from the whole quasar sample. The three redshift bins are $0.11 < z \leqslant 0.47$ (\lya{} in FUV), $0.47 < z \leqslant 1.33$ (\lya{} in NUV) and $1.33 < z \leqslant 3.5$. We recalculate $\theta$s for the three redshift ranges and present them in Fig.~\ref{fig:figsec42}. 
Color variations show similar timescale-dependent trend for all the redshift bins, indicating that they should be explained as the variation behavior of the continuum spectrum rather than the presence of emission lines. 

Particularly, in the highest redshift bin ($1.33 < z \leqslant 3.5$), FUV band probes extreme UV (EUV: 579 -- 766 $\angstrom$ at $z=1.33$) in the rest frame, and \lya{} line has been redshifted out of NUV bandpass. 
The bottom panel of Fig.~\ref{fig:figsec42} clearly demonstrates that radiation of quasar EUV, blueward of the $\sim$ 1000 \AA\ peak in the spectral energy distribution \citep{2005ApJ...619...41S}, also possesses timescale-dependent BWB trend, in consistent with the common accretion disk origin of the EUV and UV/optical radiation. 

\subsection{Comparison with the inhomogeneous accretion disk model}\label{sect:comp}
The confirmation of timescale-dependent color variability in UV bands clearly demonstrates that disk instability should play a major role. 
As introduced in Section~\ref{sect:intro}, the revised inhomogeneous disk model by Cai16 can successfully explain the timescale dependence of the color variability discovered by \citet{2014ApJ...792...54S}. 
To further explore whether the same model agrees with the observations in this work, we perform simulations of the inhomogeneous accretion disk following Cai16, and present the model predicted $\theta-\tau$ relations in Fig.~\ref{fig:figsec42} (the grey lines). 
To integrate thermal radiation of the disk, the disk is divided into square-like zones in $r$ and $\phi$ space and the logarithmic temperature of each one follows a DRW fluctuation with radius-dependent timescale (assumed to be $\tau \propto r^{3/2}$, the choice of 3/2 is hinted from the radius-dependent thermal timescale of a standard accretion disk model \citep{2001ApJ...555..775C,2009ApJ...698..895K,2012MNRAS.423..451L}). 
The simulation requires quasar parameters such as redshift, black-hole mass and Eddington ratio, which are adopted as the median values of the sample \citep{2011ApJS..194...45S}, and tabulated in Table~\ref{tab:pars}.

\begin{table}
\caption{Parameter values for the revised inhomogeneous disk model used for the three redshift bins.}
\label{tab:pars}
\centering
\begin{tabular}{lcccc}
    \hline\hline
    $z$ range & median $z$ & $\mathrm{log_{10}}(M_{\mathrm{BH}})$ & 
    $\mathrm{log_{10}}(L_{\mathrm{bol}})$ & $\mathrm{log_{10}}(\eta_{\mathrm{edd}})$ \\  
    \hline
    {(0.11,0.47]}     & 0.35   & 8.3  & 45.32  & -1.1 \\
    {(0.47,1.33]}  & 0.85   & 8.7  & 45.93  & -0.9 \\
    {(1.33,3.5]}   & 1.60   & 9.1  & 46.60  & -0.6 \\
    \hline
\end{tabular}
\end{table}

While the observed $\theta-\tau$ relation appears generally consistent with the model in the lowest redshift bin (upper panel in Fig.~\ref{fig:figsec42}), they are considerably steeper than the model predictions in the two higher redshift bins (middle and lower panels). 
To explore the reason behind this deviation, we need to look into the inhomogeneous disk model.
In the model, DRW process is invoked for variability \citep{2011ApJ...727L..24D,2016ApJ...826....7C}.
This choice is supported by the success of fitting observed light curves, structure functions and PSDs using the DRW model \citep{2009ApJ...698..895K,2010ApJ...708..927K,2010ApJ...721.1014M,2016MNRAS.459.2787K}.
As for the inhomogeneous disk model, the simulated light curves are mixed results of various DRW processes, but they can still be well fitted with a single DRW model \citep{2016ApJ...826....7C}.

Under DRW model, the structure function has a single power law slope of 0.5 at timescale ``shortward'' of the characteristic timescale (the turning point). 
The observed SDSS ensemble structure functions of quasars in Stripe 82 indeed have slopes consistent with DRW.
To be more specific, \citet{2016ApJ...826..118K} reported $0.52 \pm 0.06$ for $r$ band.
However, \GALEX{} FUV structure function appears much flatter \citep[with a fitted slope of 0.292 in FUV, ][]{2011A&A...527A..15W} than in NUV (0.439) and optical bands. 
Note their slopes was obtained by fitting the structure function in the timescale range significantly ``shortward'' of the flattening point ( $\sim$ 100 days; see Fig. 7 of Welsh et al. 2011), and thus the flatness can not be simply attributed to the smaller characteristic timescales.
And our work results in 0.52 for NUV and 0.42 for FUV. Solid deviation emerges again at least for FUV.
In line with the discrepancy between the observed $\theta-\tau$ relations and the inhomogeneous disk model based on DRW (see Fig.~\ref{fig:figsec42}), this instead suggests that the fluctuation in the inner most accretions disk, where UV radiation is produced, deviates from DRW.

At comparable timescales (from weeks to months), the X-ray power spectral density (PSD) of AGNs are known to be flatter \citep[PSD $\sim f^{-1}$, e.g.][]{2002MNRAS.332..231U,2003ApJ...593...96M,2004MNRAS.348..783M,2006Natur.444..730M,2013ApJ...770...60S} than those in the optical bands, thus can neither be described as DRW process. 
The slope $\alpha$ of the PSD is linked to the slope of SF, say $\gamma$, in the form of $\alpha = -4\gamma$ \citep[for a detailed review, please see][]{2016ApJ...826..118K}. 
This means the SF slope of EUV lies somewhere between 0.25 in X-ray and 0.5 in optical, as the EUV emission comes from the inner most accretion disk, much closer to the region where X-ray is produced (i.e., the corona). 
It is thus not surprising that its variation deviates from that of UV/optical bands, and may be closely linked with X-ray variation, which has not been considered by Cai16.

\subsection{Comparison of color variability between the two lower redshift bins}\label{sect:line}
It is interesting to note that the observed $\theta-\tau$ relation in the lowest redshift bin is flatter than those in the two higher redshift bins, and better matches the inhomogeneous disk model (Fig.~\ref{fig:figsec42}). 
This suggests that different from EUV, rest frame FUV variations could be better modeled with DRW process, similar to optical ones.
\lya{} line may also play a role as it lags behind the continuum spectrum and affects the photometry for the redshift bins differently.
In the lowest redshift bin, \lya{} line falls into the FUV bandpass. 
The \lya{} lag of NGC 5548 is known to be about 6 days \citep{2015ApJ...806..128D}.
Assuming that lag time correlates with bolometric luminosity in the form of $\tau \propto {L_{\rm bol}^{0.5}} $ \citep{2005ApJ...622..129S}, the larger bolometric luminosity  of the quasars in our sample (a median value of $10^{45.32}$ for the lowest redshift bin), in comparison with $2.6 \times 10^{44}~{\rm erg~s^{-1}}$ of NGC 5548 \citep{2015A&A...577A..37A}, results in a typical \lya{} lag of about 20 days for quasars in the lowest redshift bin. Furthermore, effect of the redshift will prolong this lag by a factor of $1+z$.
At timescales shorter than the lag, \lya{}'s emission does not tail the variation of the continuum and tend to undermine the level of color variability, thus may yield higher $\theta$ on shorter timescales, and a flatter $\theta-\tau$ relation than in higher redshift bins. 
Contrarily, \lya{} line moves into NUV band pass in the intermediate redshift bin, and could yield a steeper $\theta-\tau$ relation.

\begin{figure}[!t]
\centering
\includegraphics[width=0.45\textwidth]{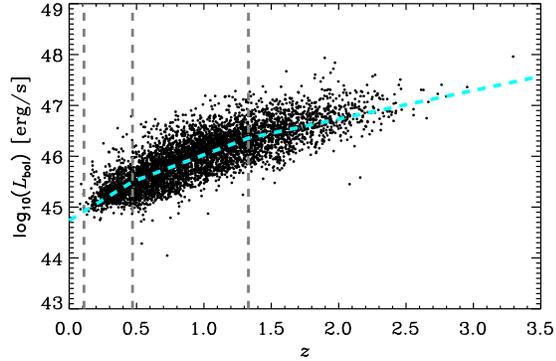}
\caption{The quasar sample's distribution on redshift vs. bolometric luminosity space. The dashed blue line indicates the dividing of brighter and dimmer subsamples for the three redshift bins, which are separated by vertical grey dashed lines.}
\label{fig:figsec51}
\end{figure}

Alternatively, in the lower redshift bin where quasars have smaller luminosities, X-ray reprocessing might be non-negligible as less luminous sources have higher X-ray to bolometric luminosity ratio \citep{2005AJ....130..387S,2010A&A...512A..34L,2010ApJS..187...64G}. Given that the color of the reprocessing radiation could be timescale independent at timescales longer than the reprocessing lag, its presence might produce a flat $\theta-\tau$ relation.
To check this possibility, within each redshift bin we divide the original sample into a brighter subsample and a dimmer one (see Fig.~\ref{fig:figsec51}). 
The division is designed to ensure that the two subsamples have consistent redshift distribution.

\begin{figure}[!t]
\centering
\includegraphics[width=0.45\textwidth]{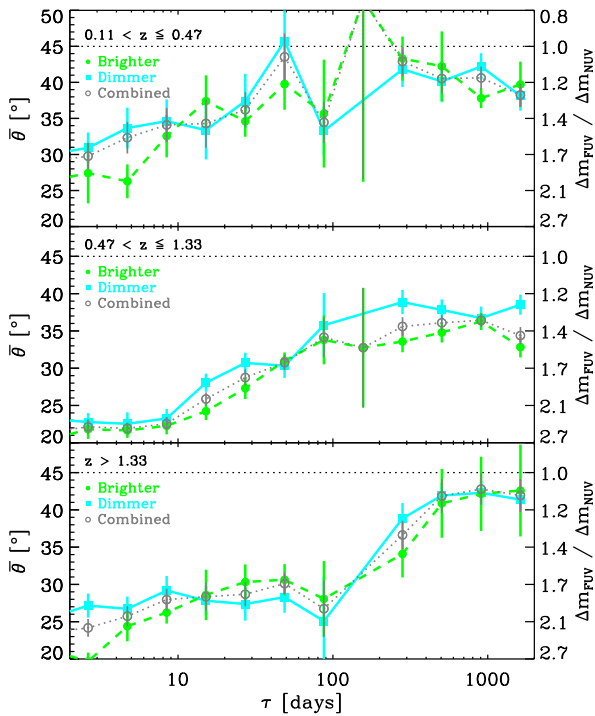}
\caption{
The color variabilities for the brighter and dimmer subsamples in three redshift bins are presented to demonstrate the dependence on the bolometric luminosity of quasars. The green dashed lines denote the brighter subsamples and the blue solid ones the dimmer quasars. The $\theta-\tau$ relations calculated from the combination of the two subsamples are presented as the thinner grey dotted lines. Both $\theta$ and the corresponding $\Delta$m$_{FUV}$/$\Delta$m$_{NUV}$ are labeled.}
\label{fig:figsec52}
\end{figure}

We present $\theta-\tau$ relations for the six subsamples in Fig.~\ref{fig:figsec52}, where within statistical uncertainties we do not see clear systematic difference between luminous and less luminous subsamples, suggesting that the flatter $\theta-\tau$ relation of quasars in the low redshift bin is unlikely simply due to their lower luminosities, or the contribution of X-ray reprocessing.

\section{Conclusions}\label{sect:conc}
Making use of FUV and NUV light curves of SDSS spectroscopically confirmed quasars collected from \GALEX{}, we confirm that the BWB trend of quasars is timescale-dependent, even in the rest frame EUV bandpass.
We show that one can use the structure functions in various bands as an indirect approach to explore the timescale dependence of color variability. 
To better understand the cause of timescale-dependent color variability, the whole sample is divided into three redshift bins and we find clear timescale dependence of the BWB trend in all three bins.
However, the observed timescale-dependent trends appear too steep to be fitted with the inhomogeneous disk model based on DRW process.
Together with the much flatter structure function observed in \GALEX{} FUV (than in NUV and optical bands), these results suggest that the inner most accretion disk, where rest frame EUV radiation is emitted, fluctuates differently from DRW.  
Studying the rest frame EUV variability, which is more or less similar to X-ray variation in PSD slopes at timescales discussed in this work, could be highly rewarding as it probes the physics in the inner most regions of the nuclei.

\section*{Acknowledgment}
We thank the referee for her/his careful reading of the manuscript and valuable comments which helps to improve this article substantially.
We are grateful for the data set used in \citet{2011A&A...527A..15W} and tips on handling them kindly provided by Jonathan M. Wheatley. 
This work is partly supported by National Basic Research Program of China (973 program, grant No. 2015CB857005) and National Science Foundation of China (grants No. 11233002, 11421303 $\&$ 11503024).
J.X.W. thanks support from Chinese Top-notch Young Talents Program, the Strategic Priority Research Program ``The Emergence of Cosmological Structures" of the Chinese Academy of Sciences (grant No. XDB09000000) and Specialized Research Fund for the Doctoral Program of Higher Education (20123402110030).
Z.Y.C. acknowledges support from the China Postdoctoral Science Foundation (grant No. 2014M560515) and the Fundamental Research Funds for the Central Universities.
\end{CJK*}
\bibliography{ref}
\end{document}